%% file: acl_latex.tex
\documentclass[11pt]{article}

\usepackage[preprint]{acl}

\usepackage{times}
\usepackage{latexsym}

\usepackage[T1]{fontenc}

\usepackage[utf8]{inputenc}

\usepackage{microtype}

\usepackage{inconsolata}

\usepackage{graphicx}

\title{A Reinforcement Learning Framework for Robust and Secure LLM Watermarking}

\author{
    \textbf{Li An\textsuperscript{1}},
    \textbf{Yujian Liu\textsuperscript{1}},
    \textbf{Yepeng Liu\textsuperscript{1}},
    \textbf{Yuheng Bu\textsuperscript{1}}\thanks{Equal advising and contribution.},
    \textbf{Yang Zhang\textsuperscript{2}${^{*}}$},
    \textbf{Shiyu Chang\textsuperscript{1}${^{*}}$}
\\
    \textsuperscript{1}UC Santa Barbara,
    \textsuperscript{2}MIT-IBM Watson AI Lab
\\
    \{\texttt{li\_an, yujianliu, yepengliu, buyuheng, chang87\}@ucsb.edu}
\\
    \texttt{Yang.Zhang2@ibm.com}
}

\usepackage{amsmath} 

\usepackage{hyperref}
\usepackage{url}
\usepackage{booktabs}
\usepackage{pifont}
\usepackage{algorithm}
\usepackage{algpseudocode}
\usepackage{wrapfig}
\usepackage{xspace}
\usepackage[table]{xcolor}
\usepackage{makecell}
\usepackage{subcaption}
\usepackage{lipsum}
\usepackage{listings}
\usepackage[most]{tcolorbox}
\usepackage{cleveref}
\usepackage{colortbl}
\usepackage{soul}
\usepackage{array}
\usepackage{enumitem}
\usepackage{multirow}
\usepackage{bm}
\usepackage{amsfonts}
\usepackage{dsfont}
\usepackage{amssymb}
\usepackage{bbm}

\usepackage{listings}
\lstdefinestyle{courierStyle}{
  basicstyle=\ttfamily,
  columns=fullflexible,
  breaklines=true,
  breakatwhitespace=true,
  postbreak=\mbox{\textcolor{gray}{$\hookrightarrow$}\space}
}
\newcommand{\e}[1]{{$#1$}}

\newcommand{\ori}{\bm x^{\text{uw}}}   
\newcommand{\wm}{\bm x^{\text{wm}}}     
\newcommand{\para}{\bm x^{\text{para-wm}}} 
\newcommand{\sent}{\bm x^{\text{sent-wm}}} 
\newcommand{\hate}{\bm x^{\text{hate-wm}}} 
\newcommand{\oripara}{\bm x^{\text{para-uw}}}  
\newcommand{\orisent}{\bm x^{\text{sent-uw}}} 
\newcommand{\orihate}{\bm x^{\text{hate-uw}}} 

\begin{document}
\maketitle

\input{Sections/0_abstract}

\input{Sections/1_intro}

\input{Sections/2_literature_review}

\input{Sections/3_preliminary}

\input{Sections/4_method}

\input{Sections/5_experiment}

\input{Sections/6_conclusion}

\clearpage
\input{Sections/limitations}
\section*{Acknowledgment}
The work of Li An, Yujian Liu, and Shiyu Chang was partially supported by National Science Foundation (NSF) Grant IIS-2338252, NSF Grant IIS-2207052, and NSF Grant IIS-2302730. The work of Yuheng Bu was partially supported by National Science Foundation (NSF) Grant OAC-2410693.

\bibliography{custom}
\clearpage
\appendix
\input{Sections/appendix}

\end{document}

%% file: Sections/0_abstract.tex
\begin{abstract}
Watermarking has emerged as a promising solution for tracing and authenticating text generated by large language models (LLMs). A common approach to LLM watermarking is to construct a green/red token list and assign higher or lower generation probabilities to the corresponding tokens, respectively. However, most existing watermarking algorithms rely on heuristic green/red token list designs, as directly optimizing the list design with techniques such as reinforcement learning (RL) comes with several challenges. First, desirable watermarking involves multiple criteria, i.e., detectability, text quality, robustness against removal attacks, and security against spoofing attacks. Directly optimizing for these criteria introduces many partially conflicting reward terms, leading to an unstable convergence process. Second, the vast action space of green/red token list choices is susceptible to reward hacking. In this paper, we propose an end-to-end RL framework for robust and secure LLM watermarking. Our approach adopts an anchoring mechanism for reward terms to ensure stable training and introduces additional regularization terms to prevent reward hacking. Experiments on standard benchmarks with two backbone LLMs show that our method achieves a state-of-the-art trade-off across all criteria, with notable improvements in resistance to spoofing attacks without degrading other criteria.
Our code is available at \url{https://github.com/UCSB-NLP-Chang/RL-watermark}.
\end{abstract}

%% file: Sections/1_intro.tex
\section{Introduction}
Large language models (LLMs) now underlie many public‑facing applications, producing text that is increasingly difficult to distinguish from human writing.  As a result, watermarking, which embeds imperceptible yet algorithmically-detectable patterns into LLM-generated text, has become a key line of defense for provenance tracking and content authentication \cite{pan2024markllm, liu2024survey, zhao2025sokwatermarkingaigeneratedcontent}.

A desired watermarking algorithm should satisfy four criteria: \ding{172} Detectability: Any watermarked text should be accurately detected, and any unwatermarked text should not be falsely detected; \ding{173} Text quality: The watermarked text should have similar quality to the unwatermarked text; \ding{174} Robustness to removal attacks: The watermarks should remain detectable under paraphrasing; and \ding{175} Security against spoofing attacks: The watermarks should be removed after malicious modifications, such as flips of sentiments and insertions of hate speech.
To design effective watermarking algorithms, a common approach is based on a green/red token list, where the token vocabulary is divided into a green list and a red list. During generation, the probability of generating green tokens is increased, and that of generating red tokens is decreased. Consequently, by counting the frequency of green versus red tokens, one can effectively detect watermarked texts~\citep{kirchenbauer2023watermark, unigram, expedit}. Although watermarking performance heavily depends on the design of the green/red list, existing approaches typically determine it randomly, leading to a suboptimal trade-off across multiple criteria.

\begin{figure}
    \centering
    \includegraphics[width=\linewidth]{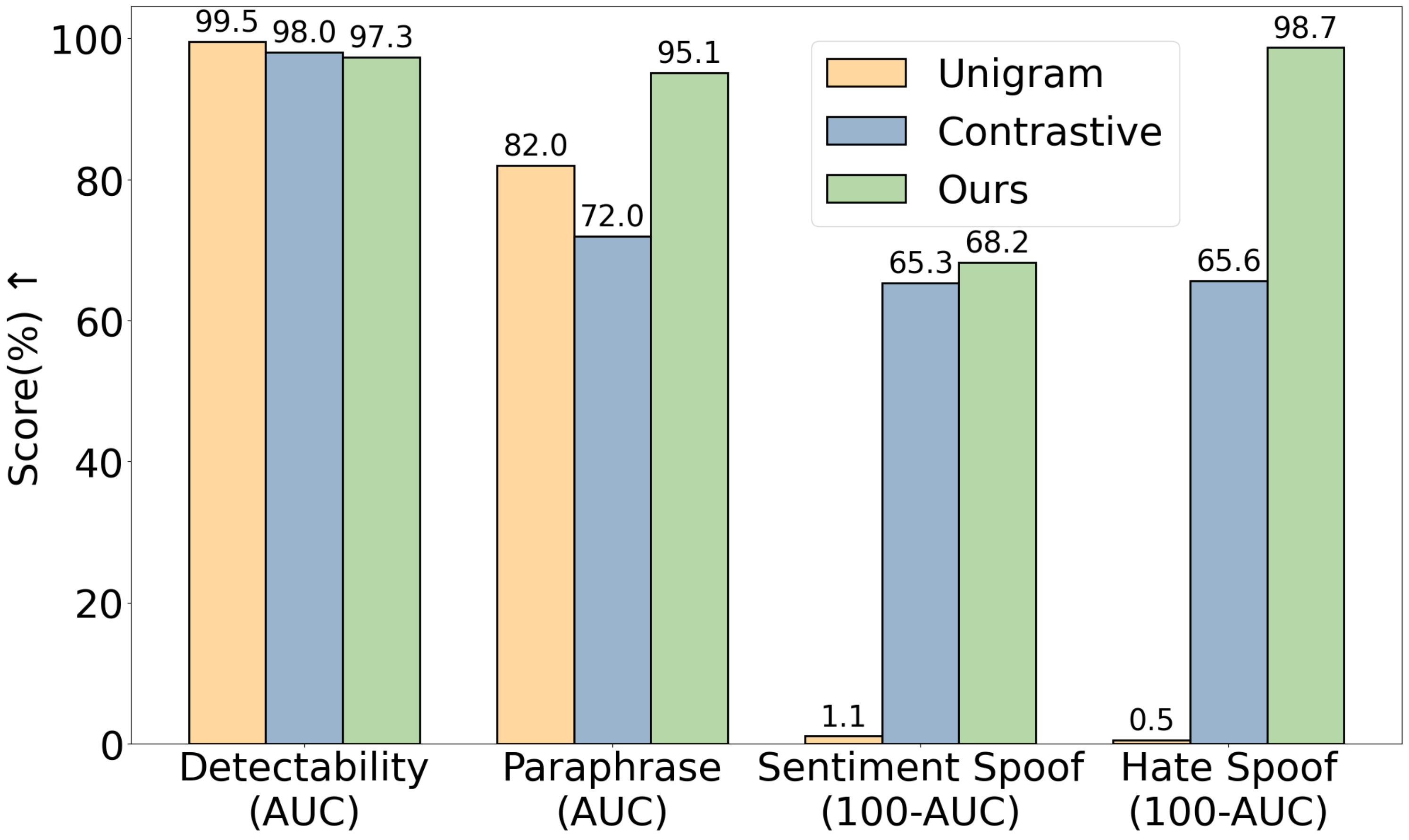}
    \vspace{-4mm}
    \caption{Performance comparison (higher is better) on detectability, robustness against paraphrase attacks, and security against sentiment and hate speech spoofing attacks. For detectability and paraphrase attack, AUCs are reported. For sentiment and hate attacks, scores are calculated using the complements (\emph{i.e.}, 100-AUC) of the corresponding AUCs.}
    \label{fig:method_comparison}
    \vspace{-2mm}
\end{figure}

More recently, semantic-aware watermarking methods~\citep{liu2024adaptive,guo2024context,liu2310semantic} have been proposed, where a mapping model encodes the prior context, and the green/red token list is determined by the semantic embeddings. By contrastively training the mapping model to be insensitive to semantic-preserving operations and sensitive to semantic-distorting operations~\citep{an2025defending}, the watermarking can be simultaneously robust to paraphrase removal attacks and secure against spoofing attacks.
However, \citet{an2025defending} only trains the model to distinguish different operations in the embedding space, which does not necessarily translate to improved end performance of watermarking. 
Figure \ref{fig:method_comparison} shows the performance of \citet{an2025defending}, which illustrates that as security improves (sentiment and hate), performance on detectability and robustness (paraphrase) degrade, indicating a tradeoff among the criteria.

In this paper, we propose an end-to-end reinforcement learning (RL) framework to directly optimize the watermarking design. Building on semantic-aware approaches, we employ a mapping model as a policy to generate the green/red token list. The policy is then optimized through a reward that balances multiple criteria under a unified framework. Specifically, given a generated green/red token list, a watermarked text is sampled, and the policy is rewarded for correctly detecting the watermark in the generated text but not in its unwatermarked counterpart. The watermarked text is then modified in two ways: through semantic-preserving edits (e.g., paraphrasing) and through semantic-distorting edits (e.g., sentiment flips and insertions of hate speech). The policy earns additional rewards when the watermark survives the first type of edits but disappears after the second.

However, such an RL training setup faces two key challenges.
The first challenge lies in the inherently conflicting nature of the watermarking criteria: detectability and robustness require watermarks to be invariant to paraphrasing, whereas security demands sensitivity to semantic-distorting modifications. Balancing these partially opposing objectives within a single reward function often leads to unstable training dynamics.
The second challenge arises from the large action space of the mapping model--each token can be assigned as a green or red token. In this setting, the model may exploit shortcuts in the reward function to achieve high reward scores without genuinely improving watermarking performance, a phenomenon analogous to reward hacking.

To address these challenges, our method includes two key designs. First, we adopt an anchored mechanism that constrains the fraction of green tokens in unwatermarked texts toward $50\%$ to address the partial conflicts among different criteria and stabilize training. Second, we augment the reward function with adversarial evaluations. In addition to applying various attacks to watermarked texts, we also generate attacks for unwatermarked texts and reward the policy if the attacked unwatermarked texts maintain a $50\%$ of green tokens. This discourages degenerate strategies such as naive hate speech detection.

We evaluate our method on two widely used benchmarks for LLM watermarking with two popular base LLMs. Experiments show that our method outperforms strong baselines by better satisfying the four criteria. Additional analyses also demonstrate the importance of our two key designs.
We summarize our contributions as follows:
\begin{itemize}[leftmargin=*, noitemsep, topsep=1pt]
\item We propose an end-to-end RL framework to optimize the green/red token list generation policy, which jointly considers detectability, robustness, and security within a unified framework.
\item We introduce an anchored reward mechanism to mitigate conflicts among reward components and stabilize training.
\item We incorporate adversarial regularization in the reward function to prevent reward hacking.
\item Experiments on standard benchmarks demonstrate that our method achieves a better trade-off among the desired criteria.
\end{itemize}

%% file: Sections/2_literature_review.tex
\section{Related Works}

\subsection{LLM Watermark}
Watermarking AI-generated text \cite{pan2024markllm, liu2024survey, zhao2025sokwatermarkingaigeneratedcontent} is crucial for ensuring transparency, accountability, and the ability to distinguish between human and machine-generated content. An in-process LLM watermark usually embeds a hidden signal directly into AI-generated text during generation by manipulating the decoding process \cite{kirchenbauer2023watermark, zhao2023provable, liu2023unforgeable, liu2023semantic, zhu2024duwak, dathathri2024scalable, lee2023wrote, he2025distributional, he2024theoretically, liu2025dataset, hu2023unbiased, christ2024undetectable, chen2025improved, kuditipudi2023robust}, fine-tuning model weights \cite{xu2024learning, zhao2023protecting, xu2025mark, block2025gaussmark}, or using a watermarking instruction \cite{liu2025context}. A post-hoc LLM watermark \cite{qiang2023natural, zhang2024remark, yang2022tracing} does not require direct access to the original LLM. Instead, it embeds a watermark into the generated text using a separate LLM, such as by paraphrasing the unwatermarked text with a watermarked LLM \cite{an2025defending} or incorporating selected keywords using LLMs \cite{chang2024postmark}. Specifically, \citet{kirchenbauer2023watermark} uses the previous token as a hash to partition the LLMs' vocabulary into green and red lists, and softly encourages the selection of green tokens during text generation to embed the watermark. \citet{guo2025optimizing} proposes a policy-driven approach for code watermarking by optimizing token selections during the next-token prediction. \citet{xu2024learning} embeds a watermark into LLM weights by co-training the LLM and detector based on RL. Our watermarking method also employs RL, but it differs significantly from \citet{xu2024learning} in two key aspects. First, \citet{xu2024learning} fully fine-tunes the LLM, whereas our approach leaves the original LLM weights untouched and instead trains a lightweight semantic mapping model, thereby maintaining the integrity of the base model. Second, rather than training a dedicated detector, we embed the watermark by perturbing the sampling process and use statistical methods for detection. 

\subsection{LLM Watermark against Spoofing Attack}
The spoofing attack \cite{sadasivan2023can, jovanovic2024watermark, pang2024no} poses a severe threat to the security of LLM watermarks, especially the reputation of the LLM owners. Specifically, \citet{sadasivan2023can} proposes a watermark forgery spoofing attack that forges watermarked text by reverse-engineering the green and red token lists used in the KGW watermarking method \cite{kirchenbauer2023watermark}. \citet{pang2024no} introduces the piggyback spoofing attack, which subtly modifies a few words to change the overall meaning or inserts harmful content, such as hate speech, into watermarked text. Although the semantic intent is significantly altered, the watermark remains detectable, potentially leading to false attribution of the harmful content to the LLM owner. To enhance the security against spoofing attacks, several semantic-aware watermarking methods have been proposed \cite{liu2024adaptive,an2025defending,yi2025unified,cai2025machine, hou2023semstamp, fu2024watermarking, ren2023robust}. Specifically, \citet{liu2024adaptive} uses a pre-trained sentence embedding model as a semantic mapping model to extract the semantic meaning of watermarked text, thereby improving resistance to forgery spoofing attacks. However, this approach struggles to defend against piggyback spoofing attacks, as pre-trained sentence embedding models often fail to capture significant semantic shifts caused by minor textual modifications. \citet{an2025defending} enhances watermark security against piggyback spoofing attacks by contrastively training a semantic mapping model to be sensitive to semantic-distorting changes while insensitive to semantic-preserving changes. This approach improves the trade-off between robustness and security. In this paper, we adopt an end-to-end training strategy to achieve a more favorable trade-off.

%% file: Sections/3_preliminary.tex
\section{Background}

\subsection{Problem Formulation}
\label{subsec:prob_formulation}
In this work, our goal is to develop a watermarking algorithm that satisfies the following four criteria:
\begin{itemize}[leftmargin=*,noitemsep,topsep=2pt]
    \item \textbf{Detectability:} The system should detect the presence of watermarks in watermarked text with a high success rate while avoiding false detection on unwatermarked text. 
    \item \textbf{Text quality:} The embedded watermarks should not disturb the quality of the generated text.
    \item \textbf{Robustness against removal attacks:} The watermarks should remain detectable after semantic-preserving edits such as paraphrasing, so that they cannot be easily removed.
    \item \textbf{Security against spoofing attacks:} The watermarks should be removed after malicious modifications like hate speech insertion and sentiment flipping, so that attackers cannot forge watermarked texts containing malicious content.
\end{itemize}
We adopt the post-hoc watermarking setting in~\citet{an2025defending}, where given an unwatermarked text, which can be generated by LLMs or humans, we generate a watermarked version of it by paraphrasing the unwatermarked input and inserting watermarks into the paraphrase.

\subsection{Semantic-aware Watermarking Algorithm}
\label{subsec:contrastive_wm}
Our method builds upon the semantic-aware watermarking algorithm~\citep{liu2024adaptive, an2025defending}. Particularly, the watermarking system comprises two components: a mapping model \e{M_{\bm \theta}} and a backbone LLM. To generate a watermarked text, the following two steps are conducted:

\noindent
\textbf{Step 1: Green/red token list construction.} 
Given an unwatermarked input text \e{\ori}, the mapping model maps it to a vocabulary-size vector \e{\bm z=M_{\bm \theta}(\ori) \in \mathbb{R}^{|\mathcal{V}|}}, where \e{\mathcal{V}} is the vocabulary.
Tokens with positive values are labeled as green tokens, forming the green token list \e{\mathcal{G}=\{v: \bm z[v] > 0\}}, while the remaining tokens constitute the red token list. Since the mapping model builds on the hidden representations of the input text, the constructed green/red token list will depend on the semantic information of \e{\ori}.

\noindent
\textbf{Step 2: Watermark injection.}
Given a paraphrasing prompt \e{\bm q} and the input text \e{\ori}, the backbone LLM generates the watermarked output \e{\wm}.
Specifically, at each decoding step \e{t}, the model produces logits \e{\bm l_t=\mathrm{LLM}(\bm q, \ori, \wm_{<t})}.
The watermark is then embedded by perturbing and increasing the logits of the green tokens: \e{\hat{\bm l}[v] = \bm l[v] \cdot (1 + \delta \mathbbm{1}(v \in \mathcal{G}))}, where \e{\delta} is the watermarking strength. Finally, the next token \e{\wm_t} is sampled according to the logits \e{\hat{\bm l}}.
In this way, the green tokens are more likely to be sampled, thereby embedding the watermark signal in generated text.

To detect the presence of watermarks in an unknown text, the same mapping model \e{M_{\bm \theta}} is used to construct the green/red token list. The text is marked as watermarked if the percentage of green tokens in the text is above a pre-defined threshold.

%% file: Sections/4_method.tex
\section{Methodology}

\begin{figure*}[t]
    \centering
    \includegraphics[width=\textwidth]{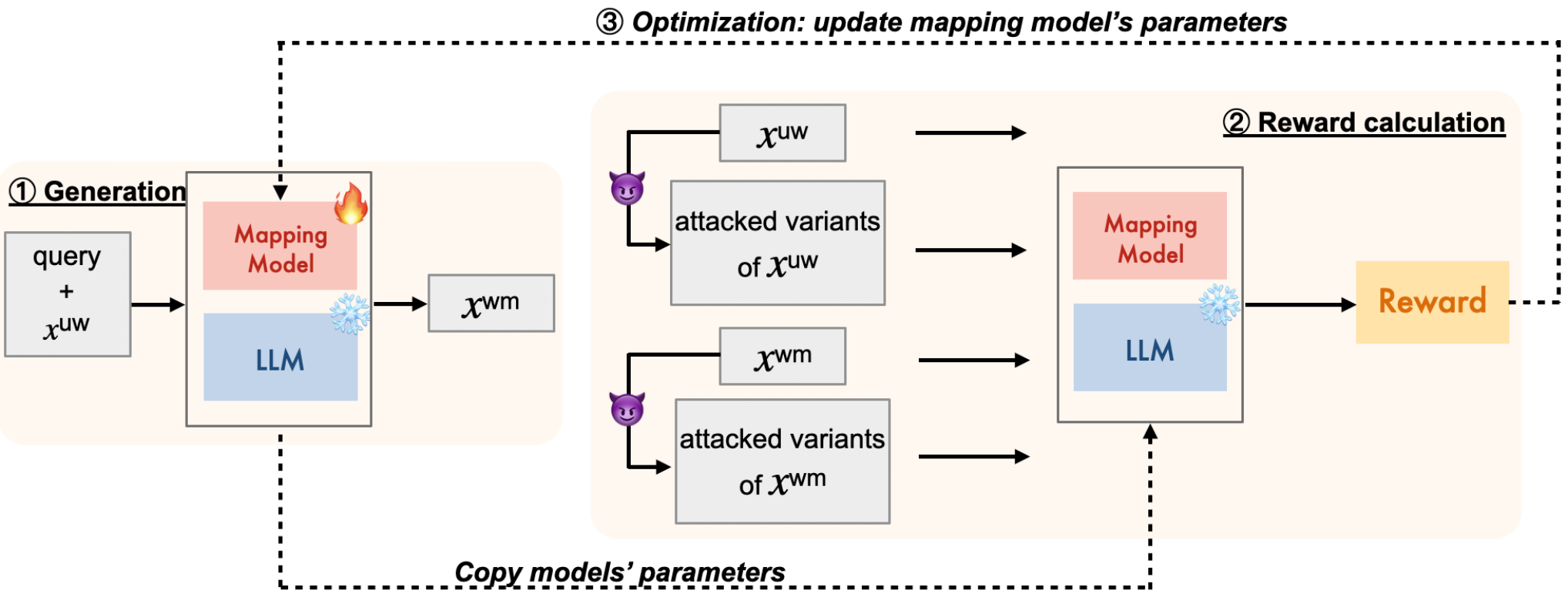}
    \vspace{-5pt}
    \caption{The RL training framework of our method. For each training instance, we perform the following three steps: \ding{172} Given an unwatermarked text \e{\ori}, sample and construct the green/red token assignment \e{\bm g}, and generate a watermarked text \e{\wm}. \ding{173} Compute the reward of \e{\bm g} by evaluating the detection score on \e{\wm} and its attacked variants, as well as on unwatermarked counterparts. \ding{174} Update the mapping model to maximize the expected reward using GRPO, while keeping the backbone LLM frozen.}
    \label{fig:framework}
\end{figure*}

In this section, we formulate the construction of the green/red token list as an RL problem and optimize the policy directly with the criteria \emph{detectability}, \emph{robustness}, and \emph{security} within a unified framework.

\subsection{RL Framework for Watermarking}
\label{subsec:rl_framework}

We cast the construction of the green/red token list as an RL problem in which the mapping model \e{M_{\bm\theta}} serves as the actor. Given an unwatermarked input \e{\ori}, the state is defined as \e{s = \ori}.
An action is a green/red assignment over the vocabulary,
\[
\bm g \in \{0,1\}^{|\mathcal{V}|}, \, \bm g[v]=1 \;\text{iff token } v \text{ is in green list}.
\]
The actor induces a stochastic policy \e{\pi_{\bm\theta}(\bm g | s)} that outputs a distribution over such assignments conditioned on the state. Taking an action corresponds to fixing the green/red token list according to \e{\bm g}. Then, a watermarked text \e{\wm} is sampled from a frozen backbone LLM following the procedure in Section~\ref{subsec:contrastive_wm}, using the green/red token list \e{\bm g}.
The environment applies a suite of attacks to \e{\wm} and returns a scalar reward \(R(s,\bm g)\) aggregating detectability, robustness, and security. The detailed reward design is in Section~\ref{subsec:reward}.

Formally, our objective is to maximize the expected reward
\vspace{-3mm}
\[
\max_{\bm\theta} \;\; \mathbb{E}_{\bm g \sim \pi_{\bm\theta}(\cdot \mid s)} \big[ R(s,\bm g) \big],
\]
which encourages policies that satisfy the desired watermarking criteria.

\paragraph{Instantiating the actor.}
To obtain a valid stochastic policy, the actor maps \e{\ori} to a vocabulary-sized vector \e{\bm z = M_{\bm\theta}(\ori)} and converts it to per-token green-list probabilities \(\bm p = \sigma(\bm z)\), where \e{\sigma(\cdot)} is the sigmoid function. An action \(\bm g\) is then sampled independently across tokens,
\[
\bm g[v] \sim \mathrm{Bernoulli}(\bm p[v]).
\]
Therefore, for a particular action \e{\bm g}, its probability can be calculated as:
\[
\pi_{\bm\theta}(\bm g | s) = \prod_{v \in \mathcal{V}} \bm p[v]^{\bm g[v]} (1 - \bm p[v])^{1 - \bm g[v]}.
\]
This stochasticity makes the list-generation process amenable to policy-gradient optimization.

\paragraph{Training procedure.}
For each training instance, we perform the following three steps, as illustrated in Figure~\ref{fig:framework}:

\noindent\textbf{Step 1: Generation.} From state \e{s=\ori}, sample \(\bm g \sim \pi_{\bm\theta}(\cdot | s)\), construct the green list \(\mathcal{G}\), and generate \(\wm\) with the frozen backbone LLM under the corresponding watermark injection.

\noindent\textbf{Step 2: Reward calculation.} Compute \(R(s,\bm g)\) by evaluating the detectability on \e{\wm} and its attacked variants, as well as on unwatermarked counterparts (details in Section~\ref{subsec:reward}).

\noindent\textbf{Step 3: Optimization.} Update \(\bm\theta\) to maximize the expected reward using GRPO~\citep{shao2024deepseekmath}, while keeping the backbone LLM frozen.

\subsection{Reward Function}
\label{subsec:reward}
We construct the reward function by jointly optimizing multiple watermarking criteria--\emph{detectability}, \emph{robustness}, and \emph{security}--within a unified formulation.
Before describing each reward component, we first define the detection score, which quantifies the likelihood that a text is watermarked.

\paragraph{Detection score.}
Given a text \e{\bm x = (x_1, \ldots, x_L)}, of length \e{L}, we obtain the per-token green-list probabilities following Section~\ref{subsec:rl_framework}: \e{\bm p=\sigma(M_{\bm \theta}(\bm x))}, where \e{\bm p[v]} represents the probability of token \e{v} being assigned as a green token. The detection score \e{D(\bm x)} is computed as the average probability of its tokens being green:\footnote{No entropy filtering~\cite{liu2024adaptive} for simplicity.}
\vspace{-4mm}
\begin{equation}
    D(\bm x) = \frac{1}{L} \sum_{t=1}^{L} \log \bm  p[x_t]. 
    \label{eq:detection_score}
\end{equation}
A higher \e{D(\bm x)} indicates a higher likelihood that the text contains watermarks.

\paragraph{Reward formulation.}
Let \e{\ori} be the original unwatermarked input, \e{\wm} its watermarked version, \e{\para} a semantic-preserving paraphrase of \e{\wm}, and \e{\sent}, \e{\hate} the spoofed variants produced by flipping the sentiments (\emph{e.g.,} positive to negative) and inserting hate speech, while preserving the main meaning of the original text (detailed procedure in Appendix~\ref{subsec:attack_details}). We instantiate three reward terms that align with the desired watermarking criteria:
\begin{itemize}[leftmargin=*,noitemsep,topsep=2pt]
    \item \textbf{Detectability:} We encourage a large margin between watermarked and unwatermarked scores, \e{D(\wm)-D(\ori)}, so that watermarks can be accurately detected.
    \item \textbf{Robustness:} We favor high detection scores on paraphrases \e{D(\para)}, so that watermarks are resilient to removal attacks.
    \item \textbf{Security:} We favor low detection scores for the spoofing-attacked versions \e{D(\sent)}, \e{D(\hate)}, so that watermarks can be removed after malicious edits.
\end{itemize}

However, naively training an RL framework with the above reward faces two key challenges.

\noindent
\textbf{Challenge 1: Partial conflicts among criteria.}
Due to the complex interplay and partial conflicts among the criteria, simply combining multiple reward terms can lead to training instability. In particular, improving detectability and robustness requires high detection scores for \e{D(\wm)}, \e{D(\para)}, and low scores for \e{D(\ori)}. At the same time, achieving strong security requires low detection scores for \e{D(\sent)} and \e{D(\hate)}, ideally lower than \e{D(\ori)}. Therefore, simply pushing down \e{D(\ori)} is insufficient, and naively summing these terms causes instability.

\noindent
\textbf{Challenge 2: Reward hacking.}
The large action space (per-token green/red assignments) enables possible shortcuts, which the model may exploit to achieve high reward scores without genuinely improving watermarking performance.
For example, to achieve low detection scores for \e{D(\hate)}, the policy could degenerate to a hate speech detector, so that as long as an input text is classified as containing hate speech, the model could assign near-zero green-list probabilities to all tokens, regardless of whether it is watermarked or not.
To address these challenges, we introduce the following two key designs in the reward function.

\noindent
\textbf{Anchored reward mechanism.}
To address the first challenge, the detection score of unwatermarked text \e{D(\ori)} should approach a neutral midpoint so that spoofed variants can be reliably driven below it, and the watermarked texts' scores can remain above it.
Therefore, we penalize deviations of the unwatermarked detection score from 0.5 by replacing the raw \e{D(\ori)} with its absolute deviation term \e{|D(\ori) - 0.5|}.

\noindent
\textbf{Anti-reward-hacking regularization.}
To discourage degenerated policies such as naive hate speech detection, we apply the same anchoring mechanism to attacked \emph{unwatermarked} texts--a paraphrase \e{\oripara}, sentiment-flipped \e{\orisent}, and hate-speech-inserted \e{\orihate}--so that the policy cannot merely learn to capture the characteristics of the attacks. This forces unwatermarked content (even after attacks) to remain neutral, preventing behaviors like hate-speech detection.

\noindent
\textbf{Final reward.}
The complete reward function combines multiple detection scores, with the regularized terms for the unwatermarked text:
\begin{equation}
\small
\begin{aligned}
R &= D(\wm)+D(\para)-D(\sent)-D(\hate) \nonumber\\
& \quad -|D(\ori)-0.5|-|D(\oripara)-0.5| \nonumber \\
& \quad -|D(\orisent)-0.5|-|D(\orihate)-0.5|.
\end{aligned}
\end{equation}

%% file: Sections/5_experiment.tex
\section{Experiments}
\input{Tables/main_result}

\subsection{Experiment Settings}
\noindent
\textbf{Evaluation datasets.}
We evaluate our method on two datasets commonly used in LLM watermarking: the \texttt{realnewslike} subset of C4~\citep{raffel2020exploring} and the LFQA dataset~\citep{krishna2023paraphrasing}. We sample 200 texts from each dataset as the original unwatermarked texts. Specifically, for C4, we use the original document, and for LFQA, we use the annotated gold completion.

\vspace{2mm}\noindent
\textbf{Metrics.}
Following \citet{an2025defending}, we report ROC-AUC scores for detecting watermarked text under four conditions: the original watermarked text, and its variants under paraphrasing, sentiment spoofing, and hate speech spoofing attacks.
Higher ROC-AUC scores on the original watermarked and paraphrased texts indicate better detectability and robustness, respectively. By contrast, lower AUC scores on the sentiment and hate speech spoofed texts are preferred, as they suggest that watermarks are successfully removed by malicious edits, reflecting stronger security.
We further compute an overall AUC score following~\citet{an2025defending}, by averaging the AUCs for detectability and robustness, along with the complements (\emph{i.e.}, 100-AUC) of the two security-related AUCs.
We also report perplexity to assess the text quality of watermarked generations.

\vspace{2mm}\noindent
\textbf{Baselines.}
We compare our approach with five baseline methods: \textsc{KGW}\citep{kirchenbauer2023watermark}, \textsc{Unigram}\citep{unigram}, \textsc{Adaptive}~\citep{liu2024adaptive}, \textsc{PostMark}~\citep{chang2024postmark}, and \textsc{Contrastive}~\cite{an2025defending}. 

The first two methods, \textsc{KGW} and \textsc{Unigram}, rely on token-level information to determine the green/red token split. \textsc{KGW} determines the green/red token split using the previous token and a random hash function, while \textsc{Unigram} improves robustness by using a fixed split. Neither method considers the semantic content of the text.
\textsc{Adaptive} introduces semantic awareness by leveraging prefix embeddings to guide the green/red split. It also utilizes the entropy of the LLM's output to adaptively choose tokens on which to inject watermarks.
\textsc{PostMark}~\citep{chang2024postmark}, a recent post-hoc method that inserts input-conditioned tokens directly into the generated text. 
Finally, we include \textsc{Contrastive}\cite{an2025defending}, which uses contrastive training to obtain a mapping model that is both sensitive to semantic-distorting operations and insensitive to semantic-preserving operations.

Although \textsc{KGW}, \textsc{Unigram}, and \textsc{Adaptive} were not originally designed for post-hoc watermarking, we adapt them to our setting by prepending a paraphrasing instruction to the target text (see Figure~\ref{fig:para-prompt} for the prompt format).
For fair comparison, we tune all methods to produce watermarked texts with similar perplexity levels, ensuring comparable text quality. The parameter details are included in Appendix \ref{subsec:baseline_param}.

\vspace{2mm}\noindent
\textbf{Training details.}
We initialize the mapping model with the contrastively trained semantic mapping model released by \citet{an2025defending} and further fine-tune it using our RL framework. We apply Group Relative Policy Optimization (GRPO) \cite{shao2024deepseekmath}, an efficient and effective RL algorithm, and adopt an unbiased gradient formulation specific to our setting (see details in Appendix \ref{subsec:ablation_reward_gradient}). For training data, we use the dataset released by \citet{an2025defending}, where only the original texts is used as unwatermarked inputs. During reward computation, we use \texttt{Qwen3-14B} \cite{qwen3technicalreport} as the attacker model to generate paraphrased and sentiment spoofed variants, and details of the attack process and prompt templates are provided in Appendix \ref{subsec:attack_details}. We evaluate the model on the validation set during training, and select the checkpoint with the highest overall AUC as the final model.

\subsection{Main Results}
Table~\ref{tab:watermark-results} compares our method with baseline watermarking algorithms across the four criteria: \emph{detectability}, \emph{robustness}, \emph{security}, and \emph{text quality}. For each backbone--\texttt{Llama-3.1-8B-Instruct} and \texttt{Qwen2.5-7B-Instruct}--we train a separate mapping model while freezing the backbone.

As can be observed, our method achieves the highest overall AUC on both datasets and backbones, while maintaining comparable perplexity to baselines. 
Traditional methods such as \textsc{KGW}, \textsc{Unigram}, \textsc{Adaptive}, and \textsc{PostMark} all struggle with removing watermarks in spoofing-attacked texts, yielding AUCs above 90 even when the text is maliciously modified. Although the \textsc{Contrastive} baseline improves spoofing resilience, the trade-off on other criteria is still suboptimal. By contrast, our end-to-end RL framework achieves a better trade-off across all three key dimensions--detectability, robustness, and security--without sacrificing perplexity. In particular, our approach not only improves robustness against paraphrasing attacks, but also shows strong security under spoofing attacks, especially hate-speech spoofing. 
The ablation study in Section~\ref{subsec:ablation} further shows how each component of our framework contributes to the overall performance improvement.

\subsection{Ablation Study}
\label{subsec:ablation}
\input{Tables/anchored_reward}

We now investigate the impacts of the two key designs in Section~\ref{subsec:reward}.

\paragraph{Anchored reward mechanism.}
To address the partial conflicts among different criteria, we introduce an anchored reward mechanism that stabilizes training by using the absolute difference between the unwatermarked text detection score and 0.5. This design prevents the model from excessively reducing or inflating the unwatermarked detection score, maintaining it near the neutral midpoint.

To evaluate the effectiveness of this design, we compare it with a naive variant of the reward function in which the anchored term is replaced by the raw unwatermarked detection score, \emph{i.e.,} \e{D(\wm) - D(\ori)}. Both models are trained under identical settings using the \texttt{Llama-3.1-8B-Instruct} backbone.
As shown in Table~\ref{tab:anchored_reward}, removing the anchored reward leads to noticeably higher AUCs on spoofing attacks, indicating worse security. Although removing the anchor brings a slight increase in detectability and robustness, the overall trade-off across all criteria is worse. This result is consistent with our observation that, without anchoring, the unwatermarked detection scores decrease excessively, making it harder to distinguish spoofing-attacked watermarked text from unwatermarked text.

\begin{figure}[h!]
    \centering
    \includegraphics[width=1.0\linewidth]{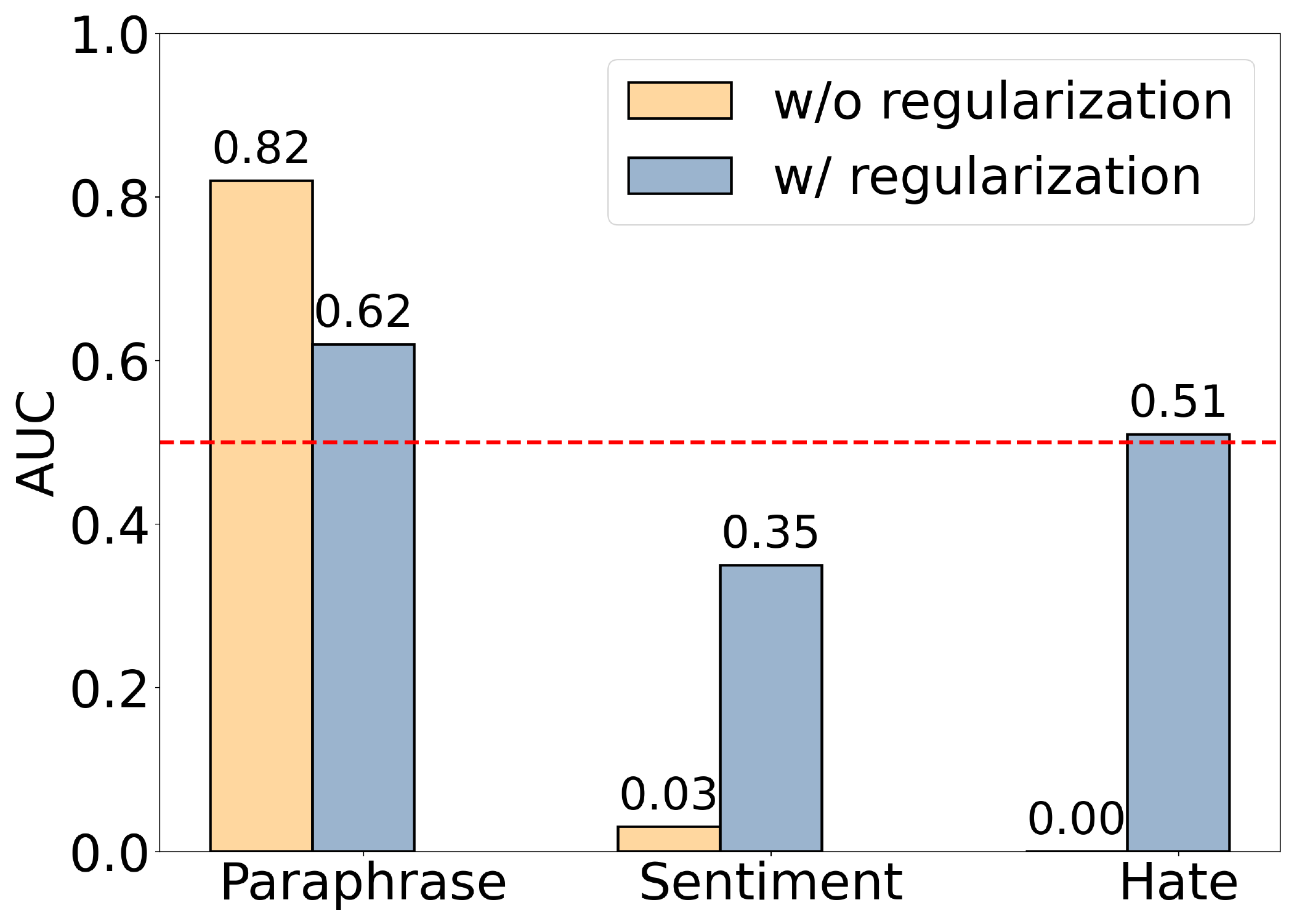}
    \caption{Effect of anti-reward-hacking regularization on AUCs of attacked unwatermarked text.}
    \label{fig:reward_hack}
    \vspace{-2mm}
\end{figure}

\paragraph{Anti-reward-hacking regularization.}
To evaluate the effectiveness of the proposed anti-reward-hacking regularization, we train a variant of the mapping model that does not contain the regularization terms, \emph{i.e.,} we only restrict the detection score of the original unwatermarked text (\e{D(\ori)}) around 0.5, but not its attacked versions (\e{D(\oripara), D(\orisent), D(\orihate)}).

For evaluation, we use the same set of 200 unwatermarked samples from the \texttt{C4} dataset as in Table~\ref{tab:watermark-results}. We apply the same paraphrasing, sentiment-spoofing, and hate-speech-spoofing prompts directly to the original unwatermarked text. We then calculate the ROC-AUC scores for distinguishing the attacked texts from the original ones.
Since the whole process does not add any watermarks, a watermarking detector should not be able to distinguish the attacked texts and original texts, so the AUC values should be close to 0.5.

As shown in Figure~\ref{fig:reward_hack}, adding regularization brings the AUCs of attacked unwatermarked variants closer to 0.5 (indicated as the red dashed line). Specifically, the model without regularization yields a high AUC for paraphrased text and near-zero AUCs for sentiment and hate-speech variants. This suggests that the model is functioning as a paraphrase, sentiment-spoofing, and hate-speech-spoofing detector, as it accurately differentiates the attacked variants from the original texts, even if no watermarks are added. By contrast, the regularized model is less accurate in identifying the attacked variants, indicating that the regularization effectively mitigates reward hacking and leads to a more reliable and interpretable reward signal.

%% file: Tables/main_result.tex
\setlength{\tabcolsep}{3pt}
\renewcommand{\arraystretch}{0.85}
\begin{table*}[t]\scriptsize
\centering
\resizebox{\textwidth}{!}{
\begin{tabular}{llccccccc}
\toprule \midrule
\multirow{4}{*}{\textbf{Method}} & \multirow{4}{*}{\textbf{Dataset}} & \multicolumn{4}{c}{\textbf{ROC-AUC (\%)}} & \multirow{4}{*}{\makecell{\textbf{Overall}\\\textbf{AUC} $\uparrow$}} & \multirow{4}{*}{\textbf{PPL}$\downarrow$} \\
\cmidrule(lr){3-6}
    &   & \multirow{3}{*}{\textbf{Detectability}} $\uparrow$ & \textbf{Robustness} $\uparrow$ & \multicolumn{2}{c}{\textbf{Security} $\downarrow$} &   &\\
\cmidrule(lr){4-6}
    &   &   &   Paraphrase & \makecell{Sentiment\\Spoof} & \makecell{Hate Speech\\Spoof} & & \\
\midrule
\rowcolor{gray!40}

\multicolumn{8}{c}{\textbf{Llama-3.1-8B-Instruct}} \\
\midrule
\multirow{2}{*}{\textsc{KGW}} & C4    & 100.00 & 72.68 & 98.85 & 100.00 & 43.46 & 8.27 \\
    & LFQA  & 100.00 & 78.03 & 99.32 & 100.00 & 44.68 & 9.04 \\
\midrule
\multirow{2}{*}{\textsc{Unigram}} & C4   & 99.54 & 81.96 & 98.44 & 99.54 & 45.88 & 8.23 \\
        & LFQA & 99.98 & 86.12 & 98.94 & 99.98 & 46.80 & 8.81 \\
\midrule
\multirow{2}{*}{\textsc{Adaptive}} & C4  & 99.78 & 72.18 & 96.50 & 99.35 & 44.03 & 8.77\\
         & LFQA  & 99.97 & 70.45 & 97.14 & 99.91 & 43.34 & 9.90 \\
\midrule
\multirow{2}{*}{\textsc{PostMark}} & C4   & 99.99 & 89.03 & 94.07 & 99.87 & 48.77 & 9.21 \\
         & LFQA & 99.93 & 87.20 & 95.54 & 99.47 & 48.03 & 9.21 \\
\midrule
\multirow{2}{*}{\textsc{Contrastive}} & C4  & 98.02	&   71.97	& 34.68	&  34.38 &  75.23 &  8.80 \\
         & LFQA  & 99.16	&  80.99	&  29.23	&    29.89 & 80.26   &  9.57 \\
\midrule
\multirow{2}{*}{\textsc{Ours}} & C4  & 97.30    & 95.11    & 31.81 & 1.31 & \textbf{89.82} & 9.01  \\
         & LFQA  &  98.08	&    100.00	&    37.31	& 3.00 &   \textbf{89.44}   &  9.83 \\
\midrule
\midrule
\rowcolor{gray!40}
\multicolumn{8}{c}{\textbf{Qwen2.5-7B-Instruct}} \\
\midrule
\multirow{2}{*}{\textsc{KGW}} & C4    & 99.12 & 67.92 & 94.04 & 99.08 & 43.48 & 8.97 \\
    & LFQA  & 99.58 & 67.38 & 95.51 & 99.56 & 42.97 & 9.23 \\
\midrule
\multirow{2}{*}{{\textsc{Unigram}}} & C4   & 97.34 & 66.99 & 93.03 & 96.13 & 43.79 & 10.03 \\
        & LFQA & 99.62 & 62.50 & 97.07 & 99.32 & 41.43 & 9.98 \\
\midrule
\multirow{2}{*}{\textsc{Adaptive}} & C4   & 99.17 & 66.08 & 91.75 & 98.49 & 43.75 & 9.77 \\
         & LFQA  & 99.26 & 61.37 & 89.96 & 98.86 & 42.95 & 10.74 \\
\midrule
\multirow{2}{*}{\textsc{PostMark}} & C4   & 99.99 & 89.03 & 94.07 & 99.87 & 48.77 & 9.21 \\
         & LFQA & 99.93 & 87.20 & 95.54 & 99.47 & 48.03 & 9.21 \\
\midrule
\multirow{2}{*}{\textsc{Contrastive}}   & C4    & 95.80 & 67.09	& 32.54	& 18.58 & 77.94 & 9.57 \\
         & LFQA  & 98.94	&  81.27   &   37.58	& 29.04 & 78.40 & 10.99 \\
\midrule
\multirow{2}{*}{\textsc{Ours}}  &   C4  &  95.12 &	98.27 &	34.16 &	3.45 & \textbf{88.95}    &   9.29    \\
         & LFQA  &   95.24 &	98.38 &	25.22 &	4.84 &  \textbf{90.89} &   11.19   \\
\midrule
\bottomrule
\end{tabular}
}
\caption{
Performance of watermarking methods. The Overall AUC is the average of the four AUC scores.
}
\label{tab:watermark-results}
\end{table*}
\setlength{\tabcolsep}{6pt}
\renewcommand{\arraystretch}{1.0}

%% file: Tables/anchored_reward.tex
\begin{table}[h!]
\centering
\small
\setlength{\tabcolsep}{1pt}
\resizebox{\linewidth}{!}{
\begin{tabular}{ccccc}
\toprule
\midrule
\multirow{3}{*}{\textbf{\makecell{Unwatermarked\\score}}} & \multicolumn{4}{c}{\textbf{ROC-AUC (\%)}} \\
                &   Detectability $\uparrow$ & Paraphrased $\uparrow$ & \makecell{Sentiment\\Spoof$\downarrow$} & \makecell{Hate Speech\\Spoof $\downarrow$} \\
\midrule
Raw   & 98.46 &   99.04   &   52.00   &   15.88  \\
Anchored  & 97.30  &   95.11   &   31.81   &   1.31  \\
\midrule
\bottomrule
\end{tabular}
}
\vspace{-1mm}
\caption{Performance of models trained using rewards with or without the anchored reward mechanism.}
\label{tab:anchored_reward}
\vspace{-1mm}
\end{table}

%% file: Sections/6_conclusion.tex
\section{Conclusion}
This paper proposes an end-to-end reinforcement learning framework for robust and secure LLM watermarking.
By directly optimizing the green–red token list generation policy under a unified reward objective, the method achieves a more balanced trade-off across multiple criteria.
An anchored reward mechanism and adversarial regularization stabilizes training and prevents reward hacking.
Experiments on multiple benchmarks and backbones show that it achieves superior resistance to removal and spoofing attacks without compromising text quality or detectability.

%% file: Sections/limitations.tex
\section*{Limitations}
Despite its promising results, the proposed approach has several limitations can be further explore in future works. First, the mapping model needs to be retrained for every new backbone LLM, which increases computational cost and limits scalability across architectures. Second, the training objective does not explicitly incorporate text quality metrics, such as fluency or coherence, which may cause the watermarking process to degrade generation quality even if detectability and robustness are improved. Finally, the experiments only consider a narrow range of attack types, mainly paraphrasing, sentiment, and hate-speech spoofing, leaving other attacks—such as translation, summarization, and learned-based stealing—unexplored.

%% file: Sections/appendix.tex
\section{Implementation Details}
\subsection{Derivation of the Unbiased Gradient}
\label{subsec:gradient}

In Section \ref{subsec:rl_framework}, we formulate the construction of the green/red token list as an RL problem. We notice that the standard policy gradient is unsufficient for our watermark setting due to the inherent duel role of the mapping model. 

Specifically,
we denote state \e{s} as the unwatermarked text \e{\ori}, and a rollout \e{\bm g}, \emph{i.e., } a way of green/red token assignment, is sampled from the policy defined by the mapping model \e{M_\theta}. The reward is defined as: \e{R_\theta(s, \bm g)}.
Our goal is to find \e{\theta} that maximizes the following objective:
\e{J(\theta) = \mathbb{E}_{\bm g \sim \pi_{\bm\theta}(\cdot \mid s)} \big[ R(s,\bm g) \big]. }
The gradient \e{\nabla_\theta J(\theta)} is:
\begin{equation}
\small
\begin{aligned}
\nabla_\theta J(\theta) &= \nabla_\theta \mathbb{E}_{\bm g \sim \pi_{\bm\theta}(\cdot \mid s)} \big[ R(s,\bm g) \big] \\
&= \nabla_\theta \int \pi_{\bm\theta}(\bm g ) R_\theta(s,\bm g )\, d\bm g  \\
&= \int R_\theta(s,\bm g ) \nabla_\theta \pi_{\bm\theta}(\bm g )\, d\bm g  + \int \pi_{\bm\theta}(\bm g ) \nabla_\theta R_\theta(s,\bm g )\, d\bm g  
\end{aligned}
\label{eq:gradient}
\end{equation}

In the last line of Eq.(\ref{eq:gradient}), the first term is the standard policy gradient.
The second term, \e{\int \pi_{\bm\theta}(\bm g ) \nabla_\theta R_\theta(s,\bm g )\, d\bm g = \mathbb{E}\left[\nabla_\theta R_\theta(s,\bm g )\right]}, is specific to our setting. The reward \e{R} is composed of multiple detection scores. And as shown in Eq.\eqref{eq:detection_score}, detection score has a non-zero gradient and thus the second term must be accounted for during optimization.

In our implementation, we take this gradient term into account. We also apply a ablation study in Appendix \ref{subsec:ablation_reward_gradient} to figure out the effectiveness of adding this gradient term. 

\subsection{Baselines}
\label{subsec:baseline_param}
To ensure a fair comparison, we tune the hyperparameters of all baseline methods such that the generated watermarked texts exhibit comparable perplexity levels, reflecting similar text quality.
Specifically, we set \e{\delta = 3.0} for \textsc{KGW} and \textsc{Unigram}; \e{\alpha = 2.0}, \e{\delta_0 = 0.1}, and \e{\delta = 0.13} for \textsc{Adaptive}; and \e{ratio = 0.06} for \textsc{PostMark}.
All other hyperparameters not listed above are kept at their default values as specified in the respective original implementations.

\subsection{RL Training Parameters}
\input{Tables/rl_param}
We optimize the parameters of the mapping model using the GRPO algorithm.
The hyperparameter settings related to RL training are summarized in Table~\ref{tab:rl_param}.
We run our experiments on four B200 GPUs.

\subsection{Attack Prompts and Detailed Implementation}
\label{subsec:attack_details}

\paragraph{Paraphrase attack.}
Given an input text, we follow the same paraphrasing prompt as in~\citet{an2025defending} to generate paraphrased variants that preserve the semantics of the original text.

\paragraph{Sentiment spoofing attack.}
The sentiment spoofing attack consists of three steps:
\ding{172} Detect the original sentiment of the input text using the prompt shown in Fig.~\ref{fig:sent-judge-prompt}.
\ding{173} Determine the target sentiment: if the detected sentiment is non-neutral, we flip it; otherwise, we randomly select either positive or negative.
\ding{174} Update the prompt shown in Fig.~\ref{fig:sent-spoof-prompt} with the target sentiment and generate the sentiment-flipped variant.

\paragraph{Hate-speech spoofing attack.}
Because modern LLMs are trained with strong safety alignment, directly prompting them to produce hate speech is challenging.
Following the procedure in~\citet{an2025defending}, we first generate a list of entity names associated with discriminatory language.
During training, to create a hate-speech–attacked variant, we prompt the LLM to produce 2–5 short hate-speech sentences containing placeholders, randomly select one entity name from the list to replace each placeholder, concatenate the sentences, and then randomly insert them into the original text.
The prompts used to generate hate-speech entities and short sentences are identical to those in~\citet{an2025defending}.

\section{Additional Results}

\subsection{Ablation Study for Reward Gradient}
\label{subsec:ablation_reward_gradient}
\input{Tables/reward_gradient}
To evaluate the effectiveness of incorporating the reward gradient into the optimization, we train two models under identical settings—one with and one without the reward gradient term.

Table~\ref{tab:reward_gradient} reports the performance comparison between the two models.
The model trained with the unbiased gradient (which explicitly includes the reward gradient term) consistently outperforms the one without it across all four criteria.
These results demonstrate that incorporating the reward gradient yields more stable optimization and leads to a better overall balance among detectability, robustness, and security.

\section{Use of AI Assistants}
AI assistant is used to assist coding and proofreading.

\input{fig/prompts}

%% file: Tables/rl_param.tex
\begin{table}[h!]
    \centering
    \begin{tabular}{lc}
    \toprule \midrule
     & \textbf{Value} \\
    \midrule
    \# of training steps & 1000 \\
    Learning rate & \e{5\mathrm{e}{-5}} \\
    Batch size & 16 \\
    \# of mini batches & 2 \\
    \# of rollouts & 8 \\
    Clipping coefficient & 0.2 \\
    Max gradient norm & 0.5 \\
    beta & 0.04 \\
    \midrule \bottomrule
    \end{tabular}
    \caption{RL training related parameters.}
    \label{tab:rl_param}
\end{table}

%% file: Tables/reward_gradient.tex
\begin{table}[h!]
\centering
\small
\setlength{\tabcolsep}{1pt}
\resizebox{\linewidth}{!}{
\begin{tabular}{ccccc}
\toprule
\midrule
\multirow{3}{*}{\textbf{\makecell{Reward\\gradient}}} & \multicolumn{4}{c}{\textbf{ROC-AUC (\%)}} \\
                &   Detectability $\uparrow$ & Paraphrased $\uparrow$ & \makecell{Sentiment\\Spoof$\downarrow$} & \makecell{Hate Speech\\Spoof $\downarrow$} \\
\midrule
w/o   & 96.35   &	78.64 &	80.07 &	59.85  \\
w/  & 97.30  &   95.11   &   31.81   &   1.31  \\
\midrule
\bottomrule
\end{tabular}
}
\caption{Performance comparison between models train using unbiased gradient or not.}
\label{tab:reward_gradient}
\end{table}

%% file: fig/prompts.tex
\begin{figure*}[htbp]
\centering
\begin{tcolorbox}
\begin{lstlisting}[style=courierStyle]
Please act as a judge and determine the sentiment of the following text. Your task is to assess whether the sentiment is positive, negative, or neutral based on the overall tone and emotion conveyed in the text. Consider factors like word choice, emotional context, and any implied feelings. The sentiment can only be chosen from 'positive', 'negative', and 'neutral'. 
Begin your evaluation by providing a short explanation for your judgment. After providing your explanation, please indicate the sentiment by strictly following this format: "[[sentiment]]", for example: "Sentiment: [[positive]]".
\end{lstlisting}
\end{tcolorbox}
\caption{Prompt used for LLM as sentiment judge.}
\label{fig:sent-judge-prompt}
\end{figure*}

\begin{figure*}[htbp]
\centering
\begin{tcolorbox}
\begin{lstlisting}[style=courierStyle]
### Task:
Modify the given text to **clearly shift its overall sentiment** to **{modified_sentiment}**. Your goal is to transform the emotional tone of the text by adjusting words, phrases, and even facts as needed. Prioritize changing emotionally charged words, evaluative adjectives, and sentiment-bearing phrases. Focus on changing or replacing specific words or short phrases to achieve the sentiment shift. Minor sentence restructuring is allowed only if it significantly helps express the new sentiment more clearly.

### Response Format:
- The LLM should explicitly state the **new sentiment of the modified text**, and provide a **brief explanation** before giving the modified text. 
- The response must strictly follow this format:

```
[MODIFIED_SENTIMENT] <modified_sentiment> [/MODIFIED_SENTIMENT]
[EXPLANATION] <modification_plan> [/EXPLANATION]
[MODIFIED_TEXT] <modified_text> [/MODIFIED_TEXT]
```
\end{lstlisting}
\end{tcolorbox}
\caption{Prompt used for sentiment spoofing attack.}
\label{fig:sent-spoof-prompt}
\end{figure*}

\begin{figure*}[htbp]
\centering
\begin{tcolorbox}
\begin{lstlisting}[style=courierStyle]
Paraphrase the following text while preserving its original meaning. Ensure that the output meets the following criteria:

1. Preserves Meaning: The paraphrase should convey the same core idea without omitting or distorting information.
2. Fluency and Grammar: The paraphrase must be natural, grammatically correct, and well-structured.
3. Appropriate Length: Maintain a similar length unless a slight adjustment improves clarity.
4. Consistency with Context: Retain the original tone and formality (e.g., academic, casual, professional).
5. Minimal Redundancy: Avoid unnecessary repetition while keeping essential details.
6. Retains Nuances: Preserve connotations, implied meanings, and idiomatic expressions where appropriate.

Just provide the paraphrased version of the text, without any introductory or concluding phrases.
\end{lstlisting}
\end{tcolorbox}
\caption{Prompt used for semantic-equivalent paraphrase.}
\label{fig:para-prompt}
\end{figure*}

%% file: acl_latex.bbl
\begin{thebibliography}{45}
\providecommand{\natexlab}[1]{#1}

\bibitem[{An et~al.(2025)An, Liu, Liu, Zhang, Bu, and Chang}]{an2025defending}
Li~An, Yujian Liu, Yepeng Liu, Yang Zhang, Yuheng Bu, and Shiyu Chang. 2025.
\newblock Defending llm watermarking against spoofing attacks with contrastive representation learning.
\newblock \emph{arXiv preprint arXiv:2504.06575}.

\bibitem[{Block et~al.(2025)Block, Sekhari, and Rakhlin}]{block2025gaussmark}
Adam Block, Ayush Sekhari, and Alexander Rakhlin. 2025.
\newblock Gaussmark: A practical approach for structural watermarking of language models.
\newblock \emph{arXiv preprint arXiv:2501.13941}.

\bibitem[{Cai et~al.(2025)Cai, Wang, Hu, and Gu}]{cai2025machine}
Yuhang Cai, Yaofei Wang, Donghui Hu, and Chen Gu. 2025.
\newblock Machine never said that: Defending spoofing attacks by diverse fragile watermark.
\newblock In \emph{The 1st Workshop on GenAI Watermarking}.

\bibitem[{Chang et~al.(2024)Chang, Krishna, Houmansadr, Wieting, and Iyyer}]{chang2024postmark}
Yapei Chang, Kalpesh Krishna, Amir Houmansadr, John Wieting, and Mohit Iyyer. 2024.
\newblock Postmark: A robust blackbox watermark for large language models.
\newblock \emph{arXiv preprint arXiv:2406.14517}.

\bibitem[{Chen et~al.(2025)Chen, Wu, Guo, and Huang}]{chen2025improved}
Ruibo Chen, Yihan Wu, Junfeng Guo, and Heng Huang. 2025.
\newblock Improved unbiased watermark for large language models.
\newblock \emph{arXiv preprint arXiv:2502.11268}.

\bibitem[{Christ et~al.(2024)Christ, Gunn, and Zamir}]{christ2024undetectable}
Miranda Christ, Sam Gunn, and Or~Zamir. 2024.
\newblock Undetectable watermarks for language models.
\newblock In \emph{The Thirty Seventh Annual Conference on Learning Theory}, pages 1125--1139. PMLR.

\bibitem[{Dathathri et~al.(2024)Dathathri, See, Ghaisas, Huang, McAdam, Welbl, Bachani, Kaskasoli, Stanforth, Matejovicova et~al.}]{dathathri2024scalable}
Sumanth Dathathri, Abigail See, Sumedh Ghaisas, Po-Sen Huang, Rob McAdam, Johannes Welbl, Vandana Bachani, Alex Kaskasoli, Robert Stanforth, Tatiana Matejovicova, and 1 others. 2024.
\newblock Scalable watermarking for identifying large language model outputs.
\newblock \emph{Nature}, 634(8035):818--823.

\bibitem[{Fu et~al.(2024)Fu, Xiong, and Dong}]{fu2024watermarking}
Yu~Fu, Deyi Xiong, and Yue Dong. 2024.
\newblock Watermarking conditional text generation for ai detection: Unveiling challenges and a semantic-aware watermark remedy.
\newblock In \emph{Proceedings of the AAAI Conference on Artificial Intelligence}, volume~38, pages 18003--18011.

\bibitem[{Guo et~al.(2024)Guo, Tian, Song, Liu, Ding, and Li}]{guo2024context}
Yuxuan Guo, Zhiliang Tian, Yiping Song, Tianlun Liu, Liang Ding, and Dongsheng Li. 2024.
\newblock Context-aware watermark with semantic balanced green-red lists for large language models.
\newblock In \emph{Proceedings of the 2024 Conference on Empirical Methods in Natural Language Processing}, pages 22633--22646.

\bibitem[{Guo et~al.(2025)Guo, Zhu, Xu, Zhang, Xiao, and Cheng}]{guo2025optimizing}
Zhimeng Guo, Huaisheng Zhu, Siyuan Xu, Hangfan Zhang, Teng Xiao, and Minhao Cheng. 2025.
\newblock Optimizing token choice for code watermarking: A rl approach.
\newblock \emph{arXiv preprint arXiv:2508.11925}.

\bibitem[{He et~al.(2024)He, Liu, Wang, Mao, and Bu}]{he2024theoretically}
Haiyun He, Yepeng Liu, Ziqiao Wang, Yongyi Mao, and Yuheng Bu. 2024.
\newblock Theoretically grounded framework for llm watermarking: A distribution-adaptive approach.
\newblock \emph{arXiv preprint arXiv:2410.02890}.

\bibitem[{He et~al.(2025)He, Liu, Wang, Mao, and Bu}]{he2025distributional}
Haiyun He, Yepeng Liu, Ziqiao Wang, Yongyi Mao, and Yuheng Bu. 2025.
\newblock Distributional information embedding: A framework for multi-bit watermarking.
\newblock \emph{arXiv preprint arXiv:2501.16558}.

\bibitem[{Hou et~al.(2023)Hou, Zhang, He, Wang, Chuang, Wang, Shen, Van~Durme, Khashabi, and Tsvetkov}]{hou2023semstamp}
Abe~Bohan Hou, Jingyu Zhang, Tianxing He, Yichen Wang, Yung-Sung Chuang, Hongwei Wang, Lingfeng Shen, Benjamin Van~Durme, Daniel Khashabi, and Yulia Tsvetkov. 2023.
\newblock Semstamp: A semantic watermark with paraphrastic robustness for text generation.
\newblock \emph{arXiv preprint arXiv:2310.03991}.

\bibitem[{Hu et~al.(2023)Hu, Chen, Wu, Wu, Zhang, and Huang}]{hu2023unbiased}
Zhengmian Hu, Lichang Chen, Xidong Wu, Yihan Wu, Hongyang Zhang, and Heng Huang. 2023.
\newblock Unbiased watermark for large language models.
\newblock \emph{arXiv preprint arXiv:2310.10669}.

\bibitem[{Jovanovi{\'c} et~al.(2024)Jovanovi{\'c}, Staab, and Vechev}]{jovanovic2024watermark}
Nikola Jovanovi{\'c}, Robin Staab, and Martin Vechev. 2024.
\newblock Watermark stealing in large language models.
\newblock \emph{arXiv preprint arXiv:2402.19361}.

\bibitem[{Kirchenbauer et~al.(2023)Kirchenbauer, Geiping, Wen, Katz, Miers, and Goldstein}]{kirchenbauer2023watermark}
John Kirchenbauer, Jonas Geiping, Yuxin Wen, Jonathan Katz, Ian Miers, and Tom Goldstein. 2023.
\newblock A watermark for large language models.
\newblock In \emph{International Conference on Machine Learning}, pages 17061--17084. PMLR.

\bibitem[{Krishna et~al.(2023)Krishna, Song, Karpinska, Wieting, and Iyyer}]{krishna2023paraphrasing}
Kalpesh Krishna, Yixiao Song, Marzena Karpinska, John Wieting, and Mohit Iyyer. 2023.
\newblock Paraphrasing evades detectors of ai-generated text, but retrieval is an effective defense.
\newblock \emph{Advances in Neural Information Processing Systems}, 36:27469--27500.

\bibitem[{Kuditipudi et~al.(2023{\natexlab{a}})Kuditipudi, Thickstun, Hashimoto, and Liang}]{expedit}
Rohith Kuditipudi, John Thickstun, Tatsunori Hashimoto, and Percy Liang. 2023{\natexlab{a}}.
\newblock Robust distortion-free watermarks for language models.
\newblock \emph{arXiv preprint arXiv:2307.15593}.

\bibitem[{Kuditipudi et~al.(2023{\natexlab{b}})Kuditipudi, Thickstun, Hashimoto, and Liang}]{kuditipudi2023robust}
Rohith Kuditipudi, John Thickstun, Tatsunori Hashimoto, and Percy Liang. 2023{\natexlab{b}}.
\newblock Robust distortion-free watermarks for language models.
\newblock \emph{arXiv preprint arXiv:2307.15593}.

\bibitem[{Lee et~al.(2023)Lee, Hong, Ahn, Hong, Lee, Yun, Shin, and Kim}]{lee2023wrote}
Taehyun Lee, Seokhee Hong, Jaewoo Ahn, Ilgee Hong, Hwaran Lee, Sangdoo Yun, Jamin Shin, and Gunhee Kim. 2023.
\newblock Who wrote this code? watermarking for code generation.
\newblock \emph{arXiv preprint arXiv:2305.15060}.

\bibitem[{Liu et~al.(2023{\natexlab{a}})Liu, Pan, Hu, Li, Wen, King, and Yu}]{liu2023unforgeable}
Aiwei Liu, Leyi Pan, Xuming Hu, Shu'ang Li, Lijie Wen, Irwin King, and Philip~S Yu. 2023{\natexlab{a}}.
\newblock An unforgeable publicly verifiable watermark for large language models.
\newblock \emph{arXiv preprint arXiv:2307.16230}.

\bibitem[{Liu et~al.()Liu, Pan, Hu, Meng, and Wen}]{liu2310semantic}
Aiwei Liu, Leyi Pan, Xuming Hu, Shiao Meng, and Lijie Wen.
\newblock A semantic invariant robust watermark for large language models, 2024.
\newblock \emph{URL https://arxiv. org/abs/2310.06356}.

\bibitem[{Liu et~al.(2023{\natexlab{b}})Liu, Pan, Hu, Meng, and Wen}]{liu2023semantic}
Aiwei Liu, Leyi Pan, Xuming Hu, Shiao Meng, and Lijie Wen. 2023{\natexlab{b}}.
\newblock A semantic invariant robust watermark for large language models.
\newblock \emph{arXiv preprint arXiv:2310.06356}.

\bibitem[{Liu et~al.(2024)Liu, Pan, Lu, Li, Hu, Zhang, Wen, King, Xiong, and Yu}]{liu2024survey}
Aiwei Liu, Leyi Pan, Yijian Lu, Jingjing Li, Xuming Hu, Xi~Zhang, Lijie Wen, Irwin King, Hui Xiong, and Philip Yu. 2024.
\newblock A survey of text watermarking in the era of large language models.
\newblock \emph{ACM Computing Surveys}, 57(2):1--36.

\bibitem[{Liu and Bu(2024)}]{liu2024adaptive}
Yepeng Liu and Yuheng Bu. 2024.
\newblock Adaptive text watermark for large language models.
\newblock \emph{arXiv preprint arXiv:2401.13927}.

\bibitem[{Liu et~al.(2025{\natexlab{a}})Liu, Zhao, Kruegel, Song, and Bu}]{liu2025context}
Yepeng Liu, Xuandong Zhao, Christopher Kruegel, Dawn Song, and Yuheng Bu. 2025{\natexlab{a}}.
\newblock In-context watermarks for large language models.
\newblock \emph{arXiv preprint arXiv:2505.16934}.

\bibitem[{Liu et~al.(2025{\natexlab{b}})Liu, Zhao, Song, and Bu}]{liu2025dataset}
Yepeng Liu, Xuandong Zhao, Dawn Song, and Yuheng Bu. 2025{\natexlab{b}}.
\newblock Dataset protection via watermarked canaries in retrieval-augmented llms.
\newblock \emph{arXiv preprint arXiv:2502.10673}.

\bibitem[{Pan et~al.(2024)Pan, Liu, He, Gao, Zhao, Lu, Zhou, Liu, Hu, Wen et~al.}]{pan2024markllm}
Leyi Pan, Aiwei Liu, Zhiwei He, Zitian Gao, Xuandong Zhao, Yijian Lu, Binglin Zhou, Shuliang Liu, Xuming Hu, Lijie Wen, and 1 others. 2024.
\newblock Markllm: An open-source toolkit for llm watermarking.
\newblock \emph{arXiv preprint arXiv:2405.10051}.

\bibitem[{Pang et~al.(2024)Pang, Hu, Zheng, and Smith}]{pang2024no}
Qi~Pang, Shengyuan Hu, Wenting Zheng, and Virginia Smith. 2024.
\newblock No free lunch in llm watermarking: Trade-offs in watermarking design choices.
\newblock \emph{Advances in Neural Information Processing Systems}, 37:138756--138788.

\bibitem[{Qiang et~al.(2023)Qiang, Zhu, Li, Zhu, Yuan, and Wu}]{qiang2023natural}
Jipeng Qiang, Shiyu Zhu, Yun Li, Yi~Zhu, Yunhao Yuan, and Xindong Wu. 2023.
\newblock Natural language watermarking via paraphraser-based lexical substitution.
\newblock \emph{Artificial Intelligence}, 317:103859.

\bibitem[{Raffel et~al.(2020)Raffel, Shazeer, Roberts, Lee, Narang, Matena, Zhou, Li, and Liu}]{raffel2020exploring}
Colin Raffel, Noam Shazeer, Adam Roberts, Katherine Lee, Sharan Narang, Michael Matena, Yanqi Zhou, Wei Li, and Peter~J Liu. 2020.
\newblock Exploring the limits of transfer learning with a unified text-to-text transformer.
\newblock \emph{Journal of machine learning research}, 21(140):1--67.

\bibitem[{Ren et~al.(2023)Ren, Xu, Liu, Cui, Wang, Yin, and Tang}]{ren2023robust}
Jie Ren, Han Xu, Yiding Liu, Yingqian Cui, Shuaiqiang Wang, Dawei Yin, and Jiliang Tang. 2023.
\newblock A robust semantics-based watermark for large language model against paraphrasing.
\newblock \emph{arXiv preprint arXiv:2311.08721}.

\bibitem[{Sadasivan et~al.(2023)Sadasivan, Kumar, Balasubramanian, Wang, and Feizi}]{sadasivan2023can}
Vinu~Sankar Sadasivan, Aounon Kumar, Sriram Balasubramanian, Wenxiao Wang, and Soheil Feizi. 2023.
\newblock Can ai-generated text be reliably detected?
\newblock \emph{arXiv preprint arXiv:2303.11156}.

\bibitem[{Shao et~al.(2024)Shao, Wang, Zhu, Xu, Song, Bi, Zhang, Zhang, Li, Wu et~al.}]{shao2024deepseekmath}
Zhihong Shao, Peiyi Wang, Qihao Zhu, Runxin Xu, Junxiao Song, Xiao Bi, Haowei Zhang, Mingchuan Zhang, YK~Li, Yang Wu, and 1 others. 2024.
\newblock Deepseekmath: Pushing the limits of mathematical reasoning in open language models.
\newblock \emph{arXiv preprint arXiv:2402.03300}.

\bibitem[{Team(2025)}]{qwen3technicalreport}
Qwen Team. 2025.
\newblock \href {https://arxiv.org/abs/2505.09388} {Qwen3 technical report}.
\newblock \emph{Preprint}, arXiv:2505.09388.

\bibitem[{Xu et~al.(2024)Xu, Yao, and Liu}]{xu2024learning}
Xiaojun Xu, Yuanshun Yao, and Yang Liu. 2024.
\newblock Learning to watermark llm-generated text via reinforcement learning.
\newblock \emph{arXiv preprint arXiv:2403.10553}.

\bibitem[{Xu et~al.(2025)Xu, Liu, Hu, Wen, and Xiong}]{xu2025mark}
Yijie Xu, Aiwei Liu, Xuming Hu, Lijie Wen, and Hui Xiong. 2025.
\newblock Mark your llm: Detecting the misuse of open-source large language models via watermarking.
\newblock \emph{arXiv preprint arXiv:2503.04636}.

\bibitem[{Yang et~al.(2022)Yang, Zhang, Chen, Zhang, Ma, Wang, and Yu}]{yang2022tracing}
Xi~Yang, Jie Zhang, Kejiang Chen, Weiming Zhang, Zehua Ma, Feng Wang, and Nenghai Yu. 2022.
\newblock Tracing text provenance via context-aware lexical substitution.
\newblock In \emph{Proceedings of the AAAI Conference on Artificial Intelligence}, volume~36, pages 11613--11621.

\bibitem[{Yi et~al.(2025)Yi, Li, Zheng, Wang, Wang, and He}]{yi2025unified}
Xin Yi, Yue Li, Shunfan Zheng, Linlin Wang, Xiaoling Wang, and Liang He. 2025.
\newblock Unified attacks to large language model watermarks: spoofing and scrubbing in unauthorized knowledge distillation.
\newblock \emph{arXiv preprint arXiv:2504.17480}.

\bibitem[{Zhang et~al.(2024)Zhang, Hussain, Neekhara, and Koushanfar}]{zhang2024remark}
Ruisi Zhang, Shehzeen~Samarah Hussain, Paarth Neekhara, and Farinaz Koushanfar. 2024.
\newblock $\{$REMARK-LLM$\}$: A robust and efficient watermarking framework for generative large language models.
\newblock In \emph{33rd USENIX Security Symposium (USENIX Security 24)}, pages 1813--1830.

\bibitem[{Zhao et~al.(2023{\natexlab{a}})Zhao, Ananth, Li, and Wang}]{unigram}
Xuandong Zhao, Prabhanjan Ananth, Lei Li, and Yu-Xiang Wang. 2023{\natexlab{a}}.
\newblock Provable robust watermarking for ai-generated text.
\newblock \emph{arXiv preprint arXiv:2306.17439}.

\bibitem[{Zhao et~al.(2023{\natexlab{b}})Zhao, Ananth, Li, and Wang}]{zhao2023provable}
Xuandong Zhao, Prabhanjan Ananth, Lei Li, and Yu-Xiang Wang. 2023{\natexlab{b}}.
\newblock Provable robust watermarking for ai-generated text.
\newblock \emph{arXiv preprint arXiv:2306.17439}.

\bibitem[{Zhao et~al.(2025)Zhao, Gunn, Christ, Fairoze, Fabrega, Carlini, Garg, Hong, Nasr, Tramer, Jha, Li, Wang, and Song}]{zhao2025sokwatermarkingaigeneratedcontent}
Xuandong Zhao, Sam Gunn, Miranda Christ, Jaiden Fairoze, Andres Fabrega, Nicholas Carlini, Sanjam Garg, Sanghyun Hong, Milad Nasr, Florian Tramer, Somesh Jha, Lei Li, Yu-Xiang Wang, and Dawn Song. 2025.
\newblock \href {https://arxiv.org/abs/2411.18479} {Sok: Watermarking for ai-generated content}.
\newblock \emph{Preprint}, arXiv:2411.18479.

\bibitem[{Zhao et~al.(2023{\natexlab{c}})Zhao, Wang, and Li}]{zhao2023protecting}
Xuandong Zhao, Yu-Xiang Wang, and Lei Li. 2023{\natexlab{c}}.
\newblock Protecting language generation models via invisible watermarking.
\newblock In \emph{International Conference on Machine Learning}, pages 42187--42199. PMLR.

\bibitem[{Zhu et~al.(2024)Zhu, Galjaard, Chen, and Chen}]{zhu2024duwak}
Chaoyi Zhu, Jeroen Galjaard, Pin-Yu Chen, and Lydia~Y Chen. 2024.
\newblock Duwak: Dual watermarks in large language models.
\newblock \emph{arXiv preprint arXiv:2403.13000}.

\end{thebibliography}
